\documentclass[twocolumn,showpacs,preprintnumbers,amsmath,amssymb,eps]{revtex4}
\usepackage{graphicx}
\usepackage{dcolumn}
\usepackage{bm}

\newcommand{\be}{\begin{equation}} \newcommand{\ee}{\end{equation}}
\newcommand{\beqn}{\begin{eqnarray}} \newcommand{\eqn}{\end{eqnarray}}

\begin{document}

\title{Measurements and analysis of the upper critical field $H_{c2}$ of
underdoped and overdoped $La_{2-x}Sr_xCuO_4$ compounds}

\author{D. H. N. Dias}
\author{E. V. L. de Mello}
\affiliation{ Instituto de F\'{\i}sica, Universidade Federal
Fluminense, Niter\'oi, R. J., 24210-340, Brazil\\}

\author{J. L. Gonzalez}
\author{H. A. Borges\\}
\affiliation{ Dept. de F\'isica, Pontif\'icia Universidade Cat\'olica de Rio de Janeiro, \\
R. Marques de S\~ao Vicente 225, R. J., 22453-900 Brasil.}
\author{A. Gomes\\}
 \affiliation{Universidade Federal do Rio de Janeiro, Rio de Janeiro, Brazil.}
\author{A. Loose\\}
\affiliation{ FU-Berlin. Berlin. Germany}

\author{A. Lopez\\}
\affiliation{ Universidade do Estado de Rio de Janeiro, Rio de Janeiro, Brazil.}

\author{F. Vieira}
\author{E. Baggio-Saitovitch\\}
\affiliation{Centro Brasileiro de Pesquisas F\'isicas, Rio de Janeiro, Brazil. }

\date{\today}

\begin{abstract}

The upper critical field $H_{c2}$ is one of the many non conventional
properties of high-$T_c$ cuprates. It is possible that the $H_{c2}(T)$ anomalies
are due to the presence of inhomogeneities in the local charge carrier
density $\rho$ of the $CuO_2$ planes.
In order to study this point, we have prepared good quality samples of
polycrystalline $La_{2-x}Sr_xCuO_{4}$  using the wet-chemical method, which has
been demonstrated to produce samples with a good cation distribution.
In particular, we have studied the temperature dependence of the upper critical
field, $H_{c2}(T)$, through magnetization  measurements on
two samples with opposite average carrier concentration ($\rho_m=x$) and nearly equal critical
temperatures, namely,
$\rho_m = 0.08$ (underdoped) and $\rho_m = 0.25$ (overdoped).
The results close to $T_c$ do not
follow the usual Ginzburg-Landau theory and are interpreted by
a theory which takes into account
the influence of the inhomogeneities.

\end{abstract}


\maketitle

\section{Introduction}

High critical temperature superconductors
(HTSC) display many non-conventional properties which remain
to be explained by a concise physical picture\cite{TS,Tallon,Lee}.
This state of affairs might be due to the fact that, differently from low
temperature superconductors, these materials have a large degree of intrinsic
inhomogeneities. Although its origin is unknown, there are many
evidences from different experiments that they  do not have
a homogeneous doping level\cite{Elbio,Mello03}. For instance, recent Scanning
Tunneling Microscopy (STM)\cite{Pan,McElroy1,McElroy2,Vershinin}
measurements have detected strong
variations in the density of states as the tip travels over a clean and sharp
surface. Also, neutron diffraction data have revealed a complex structure
of the charge distribution that has become known as "stripe structure"\cite{Tranquada}.
Nuclear Quadruple Resonance (NQR)\cite{Singer} and Angle Resolve Photo-Emission
(ARPES)\cite{DHS} have also detected inhomogeneities in the local environment
and in the charge distribution.

Based on these experimental
evidences we argue that this behavior and the inhomogeneities
are possibly due to a
phase separation transition (PST) connected with the
anomalies seen at the upper pseudogap temperature. Such PST would bring the
system to a disordered charge distribution state, with the formation of
islands or regions of distinct doping levels\cite{MelloDDias}.
Since any PST depends on the chemical mobility, this approach may
shed some light on the reason why  some compounds 
appear to be more homogeneous or, at least, do not display any gross
inhomogeneity\cite{Bobroff,Loram}, although the phase diagrams of
cuprates seem to be universal. Therefore, we think it is possible to
formulate a unified theory for the HTSC despite of their
different degrees of disorder.

In this paper we explore this possibility by showing that the anomalies
of the upper critical field $H_{c2}(T)$ as function of the
temperature $T$ are in agreement with charge carriers
inhomogeneities in two samples with completely different average
doping levels: one underdoped and other in the far overdoped
region of the superconducting phase diagram. In order to achieve
this goal this work is threefold: i) we prepared samples of
$La_{2-x}Sr_xCuO_4$ (LSCO), ii) we made several sets of $H_{c2}(T)$
measurements and iii) perform a theoretical interpretation of the
data. The non conventional features of $H_{c2}(T)$, like a positive
curvature and absence of saturation at low temperatures, are well
known from many previous experiments\cite{Ando,Wen,Berg}.
We concentrat in two samples in the underdoped and overdoped
doping levels, and measured $H_{c2}(T)$ to detect qualitative doping
dependent behavior. We interpreted the results through a theory that
takes into account the different contribution from stripes of
different local charge density. These calculations are based on the
Cahn-Hilliard (CH) theory of phase separation\cite{Otton,Mello05} for
compounds with $\rho_m\le0.20$\cite{Tallon,MelloDDias} and on a
Gaussian charge fluctuation around the average for $\rho_m\ge0.20$.
In both cases the resulting inhomogeneous systems are studied with a
Bogoliubov-deGennes approach to a disordered superconductor
\cite{MelloDDias}. Indeed there are several different approaches to
deal with the inhomogeneities in HTSC, like the method of Ghosal at
al\cite{ghosal} of a local disorder in the chemical potential,
Nunner et al\cite{Nunner} that deals with out of plane chemical
disorder, and Cabo et al\cite{Cabo} that introduced an in plane
Gaussian disorder around de average doping, to mention just a few of
what can be found in the literature. So far, all these approaches
succeeded in explain some HTSC features, what shows that the
inhomogeneities are important, but only new and refined experiments
will be able to determine the correct way to deal with them.

\section{Sample Preparation}

Several polycrystalline samples belonging to the $La_{2-x}Sr_xCuO_4$ system were prepared
by the wet-chemical method according to reference\cite{Sample,Sample2}.
Pure (99,99\%) oxide and carbonate compounds, namely $La_2O_3$, $CuO$, and
$SrCO_3$ were dried at $150^0$C and weighted with
adequate stoichiometrical proportions. The powders were dissolved into
50 ml of ultra pure acetic acid ($CH_3COOH$) and the final solution was dried
and  after heated at $900^0$C during 24 hours in flowing oxygen. After that the
powders were quenched at room temperature, then the mixtures were
reground and pressed into pellets. Finally the pre-sintered samples were sintered
at $1050^0$C during 50 hours.

\begin{figure}[!ht]
\includegraphics[width=10.0cm]{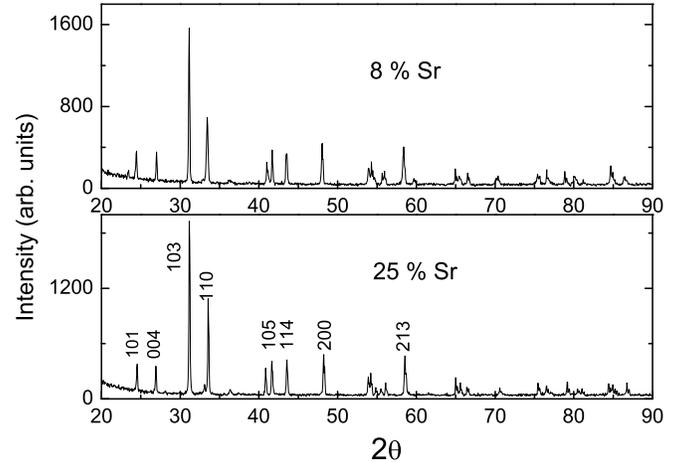}
\caption{
X-rays diffraction spectra at room temperature for both samples used in
the experiment. The principal peaks were identified according to Rietveld analysis.
} \label{raioX}
\end{figure}

Fig.(\ref{raioX}) shows the x-rays diffraction performed in both samples prepared
according the wet-chemical method. The x-ray diffraction (XRD) patterns were obtained
in a (XPert PRO PANalytical) powder diffractometer using $CuK_{\alpha}$ radiation ($\lambda=1.5418 \AA$).
Data were collected by step-scanning mode ($20^0 \le 2\theta \le 90^0$)
and 2 s counting time in
each step at room temperature. Orthorhombic (Bmab) and tetragonal (F4mmm) structure
space groups were assumed in the Rietveld analysis for the 0.08 and 0.25 strontium
percent samples respectively.

Once we have characterized the sample, we have performed magnetic
measurements by a SQUID magnetometer, as described in the next section. 


\section{Experimental Procedure}

\begin{figure}[!ht]
\includegraphics[width=10.0cm]{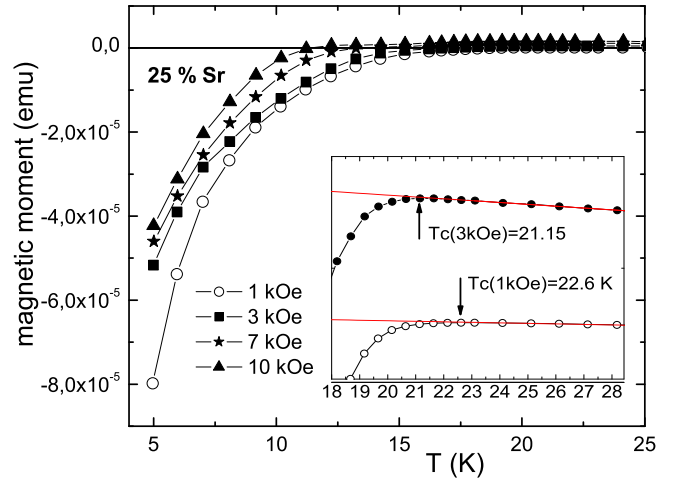}
\caption{(Color on line)  The zero-field cooling magnetic moment at
some applied magnetic fields as a function of temperature for the
sample with 25\% of strontium. The insert shows how the values
of $T_c$ were determined for two values of the applied field. }
\label{moments}
\end{figure}

\begin{figure}[!ht]
\includegraphics[width=10.0cm]{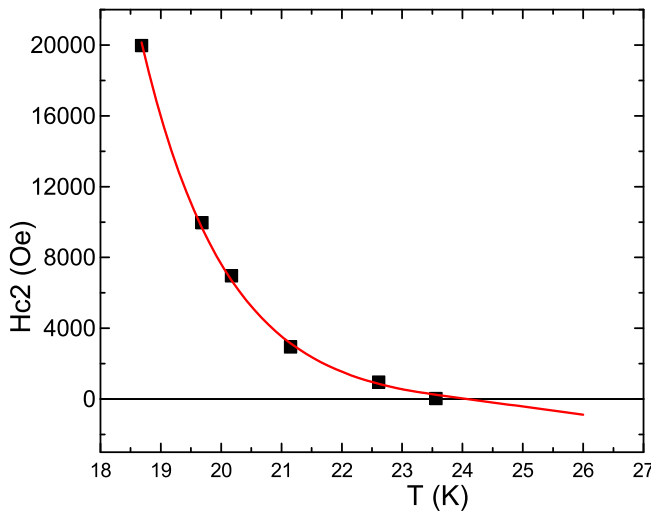}
\caption{(Color on line) Values of $T_c(H)$ calculated as shown 
in inset of Fig.(\ref{moments}) ( for $La_{1.75}Sr_{0.25}CuO_{4}$).
The value of $T_c(H_{c2}=0)$ is taken where the curve extrapolates to
zero. In this case we get $T_c(H_{c2}=0)=24.1K$, value
that will be used further on (Fig.(\ref{Hc2})).} \label{Hc2Extrap}
\end{figure}

Magnetization curves as a function of temperature were obtained with
a SQUID magnetometer in a conventional DC mode. The
measurements were performed in zero field-cooling conditions with
moderate applied fields ranging between zero and two Tesla.
Fig.(\ref{moments}) contains  sets of M(T) curves for the
sample $La_{1.75}Sr_{0.25}CuO_{4}$ (25 \% of strontium) is presented. In the
inset we show details of how $T_c(H)$ were obtained for two selected
fields, namely 3kG and 1 kG. By taking the measured values of
$T_c(H)$, we obtain the plot shown in Fig.(\ref{Hc2Extrap}), from
which we can extrapolate the value of $T_c(H_{c2}=0)$. For the
present case, we obtain $T_c(H_{c2}=0)=24.1K$ and for the underdoped
sample, we get $T_c(H_{c2}=0)=31.1K$.
A similar set of curves was obtained for the 8\% strontium sample.
In this manner we got the value of $T_c(H_{c2}=0)$ for both samples.

\begin{figure}[!ht]
\includegraphics[width=10.0cm]{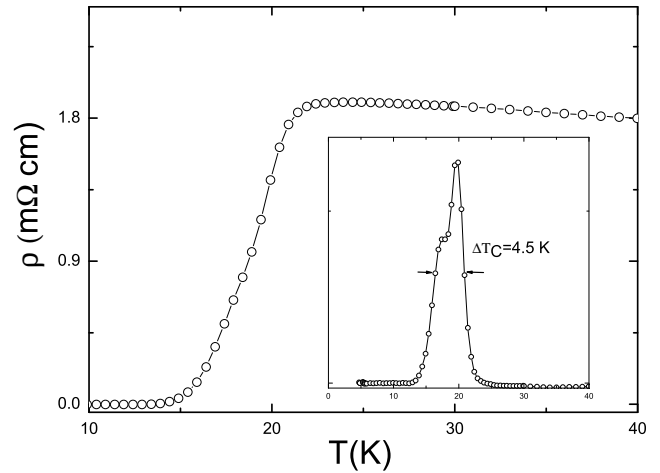}
\caption{ Resistivity as function of temperature.
The insert shows the first derivative of the data and the
maximum is taken as the superconductivity transition temperature
for this sample
at $T_c(25\%Sr)=19.9$K} \label{resisTc}
\end{figure}

On the other hand, the critical temperatures $T_c(H=0)$ for both
samples were determined from many sets of resistivity data, by
taking the maximum of the first derivative of the resistivity vs
temperature curves, namely $T_c(8\%Sr)=22.8$K and $T_c(25\%Sr)=19.9$K.
From these data, the widths of the superconducting transitions were estimated
at half-maximum of the first derivative, as displayed in Fig.(\ref{resisTc}).
For the sample with $25\%$ of strontium the $\Delta T_c$ was about 4.5 K while in
the $0.08\%$ sample the transition was wider pointing out to the presence of
strong inhomogeneities in the sample. The presence of these disorder
will be important in the discussions presented in the next sections.

\section{ PST Results and Discussion}

As discussed in the introduction,
there are many experimental evidences showing that some cuprates are
highly inhomogeneous in their charge distribution while others are not,
but all of them display the same phase diagram.
To deal with this non trivial problem, we have introduced the idea
of a PST that, depending on the mobility of the ions,
can generates various degrees of decomposition.
This phase segregation process can form
patterns on the sample, as the stripes\cite{Tranquada} or
patchwork\cite{Pan}, generating islands with different values of the
charge density, or can merely form small fluctuations
around an average doping level.
Applying a Bogoliubov-deGennes (BdG) theory to these
systems we were able to calculate the local superconducting pairing
amplitude at a given site "i" in a cluster "l".
Thus, in our calculations, a given sample with average
or mean charge density $\rho_m$ may be composed of local regions with local densities
$\rho(l)$.  Here we show
results on a $14 \times 14$ matrix, that is, 14 clusters on a stripe
form (each stripe has 14 sites) ($l=1 to 14$) and 196 sites (i runs from 1 to 196).

In general, regions with larger charge densities or doping levels
usually may become superconducting,  and those
with very low doping level are insulators and never become 
superconducting.  This anomalous
behavior is detected by the superconducting pairing amplitude
$\Delta(l,T)$ which, as the temperature is decreased,
starts at  $T_c(l)$, increases  and saturates at low temperatures.
This allows us to define a {\it  local superconducting temperature $T_c(l)$}.
%
%

\begin{figure}[!ht]
\includegraphics[width=9.0cm]{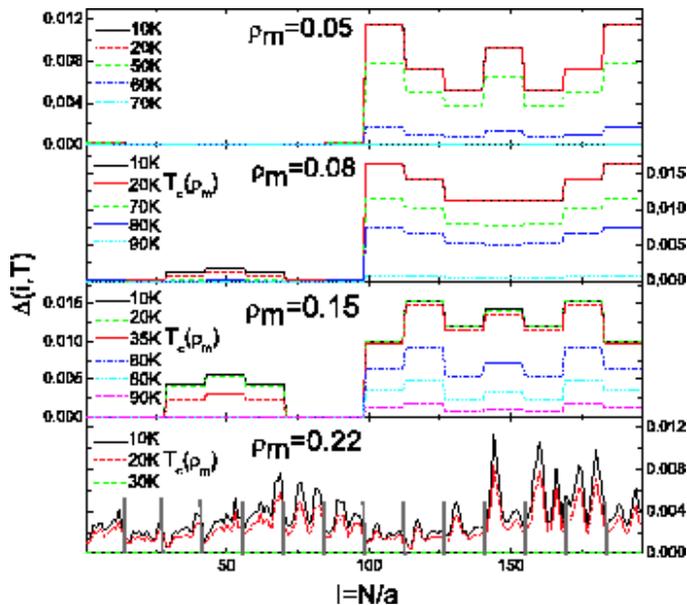}
\caption{(Color on line) Temperature evolution of the local pairing
amplitude $\Delta(i,T)$ at each stripe on a square of $14 \times
14$, that is, with 196 sites "i", for different samples. Because of
the inhomogeneities, the sites in the left have $\rho(i)\approx 0$
and the ones in the right $\rho(i)\approx 2\rho_m$. As the systems
are cooled down, more regions become superconducting and they
percolate at $T_c(\rho_m)$. These percolation threshold temperatures for
each compound are indicated in their respective panel. The panel with
$\rho=0.22$ does not have the stripe structure, because it has a 
random distribution of doping values, which are sorted following a 
Gaussian distribution.}
\label{sitios}
\end{figure}

Here we exhibit simulations
on a square mesh $14\times14$ that display stripe inhomogeneities
similar to the experimental results\cite{Tranquada}, but
derived  from a CH phase separation
theory applied to the LSCO series\cite{MelloDDias}. According to the
CH results\cite{Otton,Mello05}, the square mesh
phase separates into a bimodal distribution of charge. For
underdoped samples the phase separation is essentially total and
leads to two halves where the 7 stripes at the left are characterized by
local doping $\rho(l)\approx 0$ and the 7 at the right side have
$\rho(l)\approx 2\rho_m$ as showed in the top panel of Fig.(\ref{sitios})
(for $\rho=0.05$). As $\rho_m$ increases, the charge
fluctuations also increases, changing
the properties of a compound into metallic, and superconducting at
low temperatures. Thus, for compounds with $0.06<\rho_m<0.20$, at
low temperatures, as one can see in Fig.(\ref{sitios}), there are
also stripes with non-vanishing densities at low density regions.
Notice that the values of $\Delta$
are constant in a given stripe, in all its sites "i", and that is why
we may plot $\Delta(i,T)$ at each site, although that is meaningful only
at a given stripe $\Delta(l,T)$.
As the density of charge carriers increases, it
produces the  $\Delta(i,T)$ at the very low doping sites, i.e., in
the left region of Fig.(\ref{sitios}). When this occurs at low
temperatures, it is possible that the system
becomes superconductor at $T_c(\rho_m)$ by the percolation of all
the local superconducting regions\cite{Mello04,Mello06} and it can hold a
dissipationless current. The percolation among the local
regions where $\Delta(l,T)$ in non-zero may occur either directly
or by Josephson coupling and the associated temperature, or
the superconducting temperature  $T_c(\rho_m)$, is  also shown in
the panels of Fig.(\ref{sitios}). 
Notice that the compound with
$\rho=0.22$ does not have the stripe structure, it has just a
Gaussian distribution of doping values, since it is larger than
the PST threshold of $\rho \approx 0.2$.

\begin{figure}[!ht]
\includegraphics[width=9.0cm]{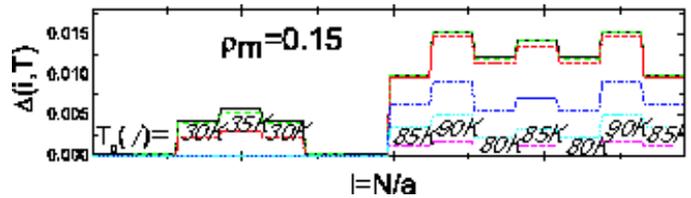}
\caption{(Color on line) To explain the concept of local
superconducting temperature and specially how certain
regions develops a non-vanishing pairing amplitude, we plot the
$\Delta(i,T)$ or $\Delta(l,T)$  for each site "i" of our two
dimensional $14 \times 14$ array. The
onset temperatures values for which these pairing amplitudes develop for each
stripe are clearly indicated in the figure, and the highest value, namely,
$T=90$K, is taken as the lower pseudogap temperature of this
compound.} \label{tci}
\end{figure}

Above $T_c(\rho_m)$, depending on the value of $\rho_m$, the compounds may be
formed by mixtures of superconducting, insulator and normal domains
and above the pairing formation temperature (the onset temperature
which sometimes is called the lower pseudogap),
they are a disordered metal with mixtures of normal ($\rho(l) >
0.05$) and insulator ($\rho(l)\le 0.05$) regions. For $\rho_m \geq
0.20$ the charge disorder is practically zero, with a small fluctuation around
$\rho_m$. From these calculations, we identify $T_{onset}(\rho_m)$
as the highest temperature ($T_c(l)$) which induces a $\Delta(l,T)$ in a
given compound which is easily seen in the panels of
Fig.(\ref{sitios}). $T_{onset}$ may be also identified with the onset of
Nernst signal\cite{MelloDDias}.
To make clear how the values of $T_{onset}$ of a given compound are
obtained, we show the larger values of $T_c(i)$ and $T_c(l)$  in
Fig.(\ref{tci}) for the sample with $\rho=0.15$.

\section {$H_{c2}$ Results and Calculations}
\begin{figure}[!ht]
\includegraphics[width=10.0cm]{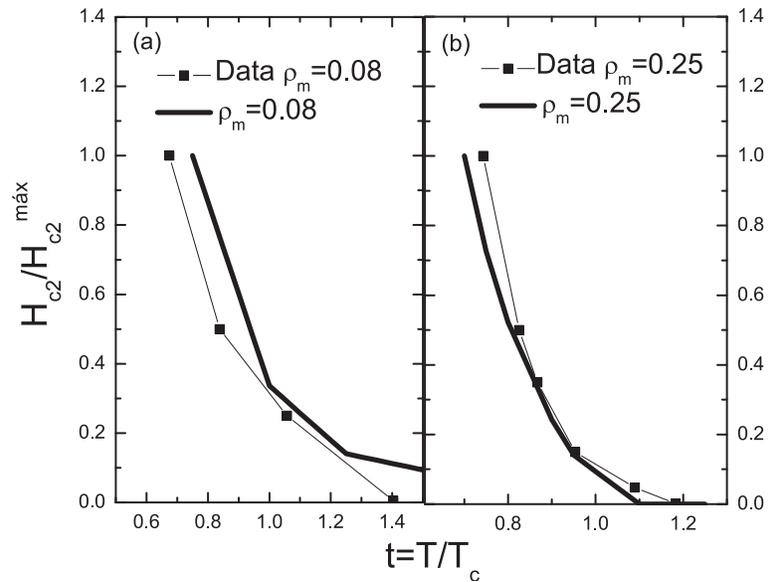}
\caption{Experimental points (connected by a thin line) against the
reduced temperature $t=T/Tc$ and the calculated curve (thick line) of
$H_{c2}$ considering inhomogeneous samples with a stripe
distribution of local superconducting temperature $T_c(l)$ for
$\rho=0.08$ and a similar calculation with a Gaussian distribution
for $\rho=0.25$. } \label{Hc2}
\end{figure}

It is well known that the Ginzburg-Landau (GL) upper critical field of a
homogeneous superconductor is a linear  function of the
temperature near $T_c$ and falls to zero at this temperature\cite{Edson}.
This behavior is not observed by our measurements, showing another
departure from conventional properties.
As we can see in  Fig.(\ref{Hc2}), the $H_{c2}(T)$ experimental points
for both samples are
linear only near and below $T/T_c \le 0.9$. As the temperature increases
it performs an upturn
curvature falling to zero further beyond $T_c$. Consequently, we need
some new ideas or theories to interpret these results.


From these $H_{c2}(T)$ curves, we can estimate that the
upper critical field goes to zero at about 31.1 K, which
is substantially higher than the value $T_c=22.8$K
of the 8 \%-Sr sample (from resistivity measurements
as discussed in section II). Similarly, $H_{c2}(T)$ falls to
zero at 24.1K for the 25 \%-Sr sample which has a $T_c=19.9$K.
Thus we see that $H_{c2}(T)$ vanishes at temperatures 8-22\%
larger than $T_c$.
This effect, the nonzero value
of $H_{c2}$ above $T_c$, is quite unusual for a normal superconductor
but it was also observed in LSCO and Bi-2201 cuprates by
the group of Wang et al\cite{Ong}. We will show
below that this result may be explained as a
consequence of the intrinsic disorder in HTSC, namely
the presence of regions with different local dopings and
distinct superconducting local temperatures $T_c(l)$.

In order to provide an interpretation to these results, we
applied a generalization of the GL $H_{c2}(T)$ expression
following along the lines described by Caixeiro et
al\cite{Edson}: The GL upper critical field near $T_c$ of a
homogeneous superconductor may be written as
\begin{eqnarray}
H_{c2}(T)&=&{\Phi_{0}\over 2\pi\xi_{ab}^2(0)}\left( {T_c-T\over
T_c}\right).\hspace{0.5cm} (T<T_c) \label{3}
\end{eqnarray}

At a temperature  $T$,
we take  each  superconducting  region in the sample characterized
by a $\rho(l)$ as the source that  generates a $\Delta(l)$ to produce a
magnetic response and displays a local $H_{c2}$,
provided that $T \le T_c(l)$. As discussed,
these regions can be in stripe or others forms, but the important
point is that they are characterized by a region of fairly constant
density $\rho(l)$ that at $T \le T_c(l)$ may shield the applied magnetic field.
Thus each of such given local region  has a local
superconducting temperature $T_c(l)$ and will contribute to the
upper critical field with a local linear upper critical field
$H_{c2}^l(T)$ near $T_c(l)$ according to the usual GL approach.
This is justified because each region has a constant density,
like a type II low temperature superconductor,
and should posses its own $H_{c2}$ that is
expected to vanish linearly at $T_c(l)$.

Consequently the total contribution of
the local superconducting regions to the  whole
sample upper critical field is the
sum of all the $H_{c2}^l(T)$'s,
\begin{eqnarray}
&&H_{c2}(T)={\Phi_{0}\over 2\pi\xi_{ab}^2(0)}{1\over W}\sum_{l=1}^W
\left( {T_c(l)-T\over T_c(l)}\right)\nonumber \\
&&={1\over W}\sum_{l=1}^W H_{c2}^l(T). \hspace{0.25cm} (T<T_c(l)\le
T_{onset}(\rho_m)) \label{4}
\end{eqnarray}
Where $W$ is the total number of superconducting regions, stripes or
islands with its local $T_c(l)\ge T $.  The maximum value
of $T_c(l)$ is the pseudogap temperature identified above
as the $T_{onset}(\rho_m)$. For the
LSCO series a coherence length of $\xi_{ab}(0)\approx 22\AA$ is used, in
accordance with the measurements\cite{Edson}. This value of
$\xi_{ab}(0)$ leads to $H_{c2}(0)$=$\Phi/2\pi\xi_{ab}^2(0)$=64T.
Due to the limitations of the GL approach, we expect the result
of this equation to be accurate only near and above the system $T_c$.

Fig.(\ref{Hc2}a) shows both the  $H_{c2}$ results of the generalized
GL calculations  together with the experimental values for
underdoped $La_{1.92}Sr_{0.08}CuO_4$ compound. In the calculations
on this compound, we used a maximum temperature of superconducting
formation, $T_{onset}\approx 90K$ (the maximum $T_c(l)$ for this
sample) from pseudogap estimates\cite{Mello03} and from our
calculations shown in Fig.(\ref{sitios}). This is the reason why the
calculated curve falls to zero at large values of the reduced
temperature $t=T/T_c$. The measured critical temperature is, by the
first derivative of the resistivity, $T_c(\rho_m)=22.8$K. Thus,
{\it using no adjustable parameters}, only values taken from
experiments, we are able to obtain very reasonable agreement with
the $H_{c2}$ experimental values and, more importantly, a clear
explanation why it does not vanish at $T_c$: at temperatures just
above $T_c$ there are some superconducting islands that do not
percolate, leading to a finite resistivity, but they are still large
enough to produce a clear magnetic response. The sum of such local
magnetic responses is clearly seen in our experiments and in other
$H_{c2}$ measurements\cite{Ong}. The magnetic contributions from non
percolated islands above $T_c$ was also measured in the form of an
anomalous magnetization\cite{Cabo,Riga} for underdoped compounds.
Such results were interpreted within the framework of the critical
state model on a charge disordered superconductor made up of
islands\cite{JL}, very close to the above approach.

Fig.(\ref{Hc2}b) also shows the data and the calculations
on the overdoped $La_{1.75}Sr_{0.25}CuO_4$. Perhaps the PST
line or upper pseudogap vanishes at $\rho_m=0.20$\cite{Tallon}, what is
in agreement with many experiments that indicate more ordered behavior
to overdoped
than the underdoped compounds\cite{MelloDDias,Bozin}.
Thus, we considered just a small Gaussian variation in the local
charge density which yielded also a small
variation in $\Delta T_c(\rho_m)\approx 18\%$. This calculation,
with small fluctuations instead of large stripe like variations,
is in agreement with the Fermi liquid behavior of overdoped samples.
Accordingly, the variations of $\Delta T_c(l)$ are very
similar  to the
compound with $\rho_m=0.22$, showed in the last panel of
Fig.(\ref{sitios}). As a consequence, the calculated $H_{c2}(T)$ curve of the
overdoped sample falls to zero just $18\%$ above $T_c$, while the
underdoped vanishes at a much larger temperature.

\section{Conclusion}

We observed that the measured $H_{c2}(T)$ curves for both underdoped
and overdoped $La_{2-x}Sr_xCuO_4$ compounds display several non-conventional
features like the positive curvature and non-vanishing values above
$T_c$.

The original GL approach to $H_{c2}$  near $T_c$ fails to
reproduce these behaviors. However it is possible to describe qualitatively
well the observed behavior by a generalization of the GL theory that
takes into account the intrinsic charge inhomogeneities. As an additional
step towards an unified description, we assumed that such disorder was originated
from a phase separation transition, possibly near the upper pseudogap
temperature. The calculations were
done in connection with the BdG formalism to obtain
the distribution of local superconducting temperature $T_c(l)$,
that is the onset of local pairing amplitude,
on  clusters with local charge density $\rho(l)$.

The measurements yield stronger non-conventional behavior for
the underdoped sample which may be an indication of a larger
degree of inhomogeneity, in agreement with other experiments\cite{McElroy2,Bozin}.
The different degrees of disorder were taken into account by the phase
separation and this unified approach  reproduced well the $H_{c2}$ results.
Thus we conclude that the observed unusual features associated with $H_{c2}$ for both samples
are consistent with the presence of charge inhomogeneities in the $La_{2-x}Sr_xCuO_4$,
which depending on the value of $x$ or $\rho_m$, appear either in the
form of stripes or in the form of small fluctuation around the average
doping level.

J. L. Gonz\'alez and A. Lopez acknowledge financial support
from Faperj. E. V. L. de Mello is grateful to CNPq for partial
financial support.


%


\begin{references}
%
%
\bibitem{TS} T. Timusk and B. Statt. Rep. Prog. Phys. \textbf{62}, 61 (1999).
\bibitem{Tallon}J. L. Tallon and J. W. Loram, Physica C \textbf{349}, 53 (2001).
%
%
\bibitem{Lee} Patrick A. Lee, Naoto Nagaosa, Xiao-Gang Wen, Rev. Mod. Phys. \textbf{78}, 17 (2006).
%
\bibitem{Elbio} Elbio Dagotto, Science {\bf 309}, 257 (2005), and
{\it Nanoscale Phase Separation and Colossal Magnetoresistence}, Springer
(2002).
%
\bibitem{Mello03}  E.V.L. de Mello, E.S. Caixeiro, and
J.L. Gonz\'alez,  Phys. Rev. {\bf B67}, 024502 (2003).
%
\bibitem{Pan}
S. H. Pan et al., Nature, {\bf 413}, 282-285 (2001) and
cond-mat/0107347.
 %
\bibitem{McElroy1} K. McElroy, R.W Simmonds, J. E. Hoffman, D.-H. Lee,
J. Orenstein, H. Eisaki, S. Uchida, J.C. Davis, Nature,
{\bf 422}, 520 (2003).
%
\bibitem{McElroy2} K. McElroy, D.-H. Lee, J. E. Hoffman, K.M. Lang,
J. lee, E. W. Hudson, H. Eisaki, S. Uchida,  J.C. Davis, Phys.
Rev. Lett., {\bf 94}, 197005 (2005).
%
\bibitem{Vershinin}  Michael Vershinin, Shashank Misra, S. Ono, Y. Abe,
Yoichi Ando, Ali Yazdani, Science {\bf 303}, 1995 (2004).
%
\bibitem{Tranquada}
J.M.Tranquada, B.J. Sternlieb, J.D. Axe, Y. Nakamura,
and S. Uchida, Nature (London),375, 561 (1995).
%
\bibitem{Singer} P. M. Singer, A. W. Hunt, and T. Imai, Phys. Rev.
 Lett. {\bf 88}, 047602 (2002).
%
\bibitem{DHS}  A. Damascelli, Z. Hussain, and Z.-X. Shen, Rev. Mod. Phys.
{\bf75}, 473 (2003)
%
\bibitem{MelloDDias}
E. V. L. de Mello and D. N. Dias, J. Phys.: Condens. Matter {\bf 19}
(2007) 086218
%
\bibitem{Bobroff}
J. Bobroff, H. Alloul, S. Ouazi, P. Mendels, A. Mahajan,
N. Blanchard, G. Collin, V. Guillen, and J.-F. Marucco,
Phys. Rev. Lett. {\bf 89} 157002 (2002).
%
\bibitem{Loram}
J.W. Loram, J.L. Tallon, and W. Y. Liang, Phys. Rev.
{\bf B} 69, 060502 (2004).
%
\bibitem{Ando}
Y. Ando, G.S. Boebinger, A. Passner, L.F. Schneemeyer, T. Kimura, M. Okuya, S. Watauchi,
J. Shimoyama, K. Kishio, K. Tamasaku, N. Ichikawa, and S. Uchida,
Phys. Rev. B {\bf 60}, 12475 (1999).
%
\bibitem{Wen} H. H. Wen, X. H. Chen, W. L. Yang, and Z. X. Zhao,
Phys. Rev. Lett. {\bf 85}, 2805 (2000).
%
\bibitem{Berg}
H. Berg, R. Muller, R. Borowski, B. Freitag, I. Muller and B. Roden,
Journal of Alloys and Compounds {\bf 267}, 279 (1998).
%
\bibitem{Otton}
E.V.L de Mello, and Otton T. Silveira Filho Physica {\bf A} 347, 429 (2005).
%
\bibitem{Mello05} E.V.L de Mello, and E. S. Caixeiro, J. Supercond.  {\bf 18}, 653 (2005).
%
\bibitem{ghosal}
A. Ghosal et. al., Phys. Rev. B {\bf 65}, 014501 (2001).
%
\bibitem{Nunner}
T.S. Nunner, B.M. Andersen, A. Melikyan, and P.J. Hirschfeld,
Phys. Rev. Lett. {\bf 95}, 177003 (2005).
%
\bibitem{Cabo} Lucia Cabo et al, Phys. Rev. {\bf B73}, 184520 (2006).
%
\bibitem{Sample} T. Nagano, Y. Tomioka, Y. Nakayama, K. Kishio and K. Kitazawa,
Physical Review {\bf B 48}, 9689 (1993).
%
\bibitem{Sample2} H. Berg, R. Muller, R. Borowski, B. Freitag, I. Muller and B. Roden,
Journal of Alloys and Compounds {\bf 267}, 279 (1998).
%
\bibitem{Mello04} E.V.L. de Mello, and E.S. Caixeiro, Phys. Rev. B {\bf 70},
224517 (2004).
%
\bibitem{Mello06} E.V.L de Mello, and  E.S. Caixeiro, J. Phys. Chem. Sol.
{\bf 67}, 165 (2006).
%
\bibitem{Edson}
E. S. Caixeiro, J. L. Gonz\'alez, and E. V. L. de Mello, Phys. Rev.
{\bf B69}, 024521 (2004),
and Physica {\bf 408-410}, 348 (2004).
%
\bibitem{Ong} Yayu Wang, Lu Li, and N. P. Ong, Phys. Rev. B {\bf 73},
024510 (2006).
%
\bibitem{Riga} A. Lascialfari et al, Phys. Rev. {\bf B 68}, 100505(R) (2003).
%
%
%
\bibitem{JL} J. L. Gonz\'alez, and E. V. L. de Mello, Phys. Rev. {\bf B 69}, 134510 (2004)
and Physica {\bf 408-410}, 441 (2004).
%
%
%
\bibitem{Bozin} E.S. Bozin, G.H. Kwei, H. Takagi, and S.J.L. Billinge,
Phys. Rev. Lett. {\bf 84}, 5856, (2000).
%

\end{references}
\end{document}